\documentclass[onecolumn, a4size, 11pt]{IEEEtran}
\usepackage{amsmath}
\usepackage{amssymb}
\usepackage{amsfonts}
\usepackage{graphicx}
\usepackage{epsfig}
\usepackage{subfigure}
\usepackage{psfrag}

\title{Optimal Dynamic Resource Allocation for Multi-Antenna Broadcasting with Heterogeneous Delay-Constrained Traffic
\footnote{Submitted to
IEEE Journal on Selected Topics in Signal Processing (JSTSP),
special issue on MIMO-optimized transmission systems for delivering
data and rich content, July 15, 2007, revised February 28, 2008. Rui
Zhang is with the Institute for Infocomm Research, Singapore.
E-mail: rzhang@i2r.a-star.edu.sg}}

\author{Rui Zhang}

\begin{document}
\maketitle \maketitle \thispagestyle{empty}

\begin{abstract}
This paper is concerned with dynamic resource allocation in a
cellular wireless network with slow fading for support of data
traffic having heterogeneous transmission delay requirements. The
multiple-input single-output (MISO) fading broadcast channel (BC) is
of interest where the base station (BS) employs multiple transmit
antennas to realize simultaneous downlink transmission at the same
frequency to multiple mobile users each having a single receive
antenna. An information-theoretic approach is taken for
characterizing capacity limits of the fading MISO-BC under various
transmission delay considerations. First, this paper studies
transmit optimization at the BS when some users have delay-tolerant
``packet'' data and the others have delay-sensitive ``circuit'' data
for transmission at the same time. Based on the convex optimization
framework, an online resource allocation algorithm is derived that
is amenable to efficient cross-layer implementation of both physical
(PHY) -layer multi-antenna transmission and media-access-control
(MAC) -layer multiuser rate scheduling. Secondly, this paper
investigates the fundamental throughput-delay tradeoff for
transmission over the fading MISO-BC. By comparing the network
throughput under completely relaxed versus strictly zero
transmission delay constraint, this paper characterizes the limiting
loss in sum capacity due to the vanishing delay tolerance, termed
the {\it delay penalty}, under some prescribed user fairness for
transmit rate allocation.
\end{abstract}

\vspace{0.1in}

\begin{keywords}
Broadcast channel (BC), fading channel, multi-antenna,
throughput-delay tradeoff, dynamic resource allocation, cross-layer
optimization, convex optimization.
\end{keywords}

\setlength{\baselineskip}{1.3\baselineskip}
\newtheorem{claim}{Claim}
\newtheorem{guess}{Conjecture}
\newtheorem{definition}{Definition}[section]
\newtheorem{fact}{Fact}
\newtheorem{assumption}{Assumption}
\newtheorem{theorem}{Theorem}[section]
\newtheorem{lemma}{Lemma}[section]
\newtheorem{ctheorem}{Corrected Theorem}
\newtheorem{corollary}{Corollary}
\newtheorem{proposition}{Proposition}
\newtheorem{example}{Example}
\newtheorem{remark}{Remark}[section]
\newtheorem{problem}{Problem}[section]
\newtheorem{algorithm}{Algorithm}[section]
\newcommand{\mv}[1]{\mbox{\boldmath{$ #1 $}}}

\section{Introduction}

In mobile wireless networks, communications typically take place
over time-varying channels. When this time-variation or fading is
``fast'' such that the channel state information (CSI) is hardly
obtainable at the transmitter, a classical approach for mitigating
impairments of fading to transmission reliability is to apply {\it
diversity techniques}, such as coded diversity, antenna diversity,
and path diversity. On the other hand, when the fading channel
changes sufficiently ``slowly'' such that the transmitter is able to
acquire the CSI, a general approach to compensate for the fading is
{\it dynamic resource allocation}, whereby transmit resources such
as power, bit-rate, antenna-beam and bandwidth are dynamically
allocated based upon the fading distribution. Effective
implementation of dynamic resource allocation usually requires joint
optimization of both physical (PHY) -layer transmission and
media-access-control (MAC) -layer rate scheduling in the classical
communication protocol stack, and thus demands for a new {\it
cross-layer} design methodology.

One challenging issue to be addressed for dynamic resource
allocation in tomorrow's wireless networks is how to meet with
user's heterogeneous transmission quality-of-service (QoS)
requirements. Among others, the demand for wireless high-speed
connectivity for both delay-tolerant ``packet'' data and
delay-sensitive ``circuit'' data is expected to rise significantly
in the next decade. Therefore, study on both spectral and power
efficient transmission schemes for support of {\it heterogeneous
delay-constrained data traffic} becomes an important area for
research. On the other hand, because tolerance for a larger
transmission delay incurred to data traffic allows for more flexible
transmit power and rate adaptation over time and thereby leads to a
larger transmission throughput in the long term, there is in general
a fundamental {\it throughput-delay tradeoff} associated with
dynamic resource allocation over fading channels. Characterization
of such fundamental tradeoff is another important research problem
because it reveals the ultimate gain achievable by dynamic resource
allocation under realistic transmission delay requirements.

This paper is aimed to provide concrete answers to the
aforementioned problems by considering the fading broadcast channel
(BC) that models the downlink transmission in a typical wireless
cellular network. An information-theoretic approach is taken in this
paper to address some fundamental limits of dynamic resource
allocation for the fading BC under various transmission delay
considerations. In particular, the fading multiple-input
single-output (MISO) BC is considered where multi-antennas are
equipped at the transmitter of the base station (BS), and single
antenna at the receiver of each mobile user. Because of
multi-antennas at the transmitter, spatial multiplexing can be used
at the BS to support simultaneous transmission to mobile users at
the same frequency, named space-division-multiple-access (SDMA). A
slow-fading environment is assumed, and for simplicity, the
block-fading (BF) channel model is adopted. It is further assumed
that the BS has perfect user CSI at its transmitter, and is thus
able to perform a centralized dynamic resource allocation based upon
multiuser channel conditions. This paper's main contributions are
summarized as follows:
\begin{itemize}

\item This paper studies optimal dynamic resource allocation for the fading MISO-BC when both
no-delay-constrained (NDC) packet data and delay-constrained (DC)
circuit data are required for transmission at the same time. A
cross-layer optimization approach is taken for jointly optimizing
capacity-achieving multi-antenna transmission at the PHY-layer and
fairness-ensured multiuser rate scheduling at the MAC-layer. A
convex optimization framework is formulated for minimizing the
average transmit power at the BS subject to both NDC and DC user
rate constraints. A {\it two-layer Lagrange-duality method} is shown
to be the key for solving this problem. Based on this method, a
novel online resource allocation algorithm that is amenable to
efficient cross-layer implementation is derived, and its convergence
behavior is validated.

\item This paper investigates the fundamental throughput-delay
tradeoff for the fading MISO-BC under optimal dynamic resource
allocation. By taking the difference between the maximum sum-rate of
users under NDC and DC transmission subject to the constraint that
the rate portion allocated to each user needs to be regulated by the
same prescribed {\it rate-profile}, the paper presents a novel
characterization for the limiting loss in sum capacity due to the
vanishing delay tolerance, termed the {\it delay penalty}, for the
fading MISO-BC. Thereby, the delay penalty provides the answer to
the following interesting question: Comparing no delay constraint
versus zero-delay constraint for all users in the network, how much
is the maximum percentage of throughput gain achievable for {\it
all} users by optimal dynamic resource allocation?
\end{itemize}

The capacity region under NDC or DC transmission for a fading
single-input single-output (SISO) BC has been characterized in
\cite{Li01a}, \cite{Li01b}, and for a fading SISO multiple-access
channel (MAC) in \cite{Tse98a}, \cite{Tse98b}. A similar scenario
like in this paper with mixed NDC and DC transmission has also been
considered in \cite{Liang06} for the single-user multiple-input
multiple-output (MIMO) fading channel, and in \cite{Jindal03} for
the fading SISO-BC. The comparison of achievable rates between NDC
and DC transmission has been considered in \cite{Huang03} for the
fading MIMO-BC. However, none of the above prior work has considered
transmit optimization with mixed NDC and DC data traffic for the
fading MISO-BC, which is addressed in this paper. On the other hand,
throughput-delay and power-delay tradeoffs for communications over
fading channels by exploiting the combined CSI and data buffer
occupancy at the transmitter have been intensively studied in the
literature for both single-user and multiuser transmission (e.g.,
\cite{Berry04} and references therein). In contrast to prior work,
this paper studies the throughput-delay tradeoff from a new
perspective by characterizing the fundamental delay penalty in the
network throughput owing to stringent (zero) transmission delay
constraint imposed by all users. The concept of rate-profile, or its
equivalent definitions for specifying some certain fairness in user
rate allocation have also been considered for the SISO multiuser
channel in \cite{BalancedCapacity}, \cite{Shen05}, and for the MIMO
multiuser channel in \cite{Mohseni06}, \cite{TJ07}. However, to the
author's best knowledge, application of rate-profile for
characterizing the delay penalty in a multiuser fading channel is a
novelty of this paper. There has been recently a great deal of study
on the real-time resource allocation algorithm, named proportional
fair scheduling (PFS) (e.g., \cite{Tse02}-\cite{Caire07} and
references therein), which maximizes the network throughput by
exploiting the multiuser channel variation and at the same time,
maintains certain fairness among users in rate allocation. However,
PFS is unable to guarantee any prescribed user rate demand. In this
paper, a novel online scheduling algorithm is proposed to ensure
that all NDC and DC user rate demands are satisfied with the minimum
transmit power consumption at the BS.

The remainder of this paper is organized as follows. Section
\ref{sec:system model} illustrates the fading MISO-BC model and
provides a summary of known information-theoretic results for it.
Section \ref{sec:mixed traffic} addresses the optimal cross-layer
dynamic resource allocation problem for support of simultaneous
transmission of heterogeneous delay-constrained traffic. Section
\ref{sec:tradeoffs} characterizes the fundamental throughput-delay
tradeoff for the fading MISO-BC. Section \ref{sec:simulation
results} provides the simulation results. Finally, Section
\ref{sec:conclusion} concludes the paper.

{\it Notation}: This paper uses upper case boldface letters to
denote matrices and lower case boldface letters to indicate vectors.
For a square matrix $\mv{S}$, $|\mv{S}|$ and $\mv{S}^{-1}$ are its
determinant and inverse, respectively. For any general matrix
$\mv{M}$, $\mv{M}^{\dag}$ denotes its conjugate transpose. $\mv{I}$
and $\mv{0}$ indicate the identity matrix and the vector with all
zero elements, respectively. $\|\mv{x}\|$ denotes the Euclidean norm
of a vector $\mv{x}$. $\mathbb{E}_n[\cdot]$ denotes statistical
expectation over the random variable $n$. $\mathbb{R}^M$ denotes the
$M$-dimensional real Euclidean space and $\mathbb{R}^M_{+}$ is its
non-negative orthant. $\mathbb{C}^{x \times y}$ is the space of
$x\times y$ matrices with complex number entries. The distribution
of a circularly-symmetric complex Gaussian (CSCG) vector with the
mean vector $\mv{x}$ and the covariance matrix $\mv{\Sigma}$ is
denoted by $\mathcal{CN}(\mv{x},\mv{\Sigma})$, and $\sim$ means
``distributed as''. $\{x\}^+$ denotes the non-negative part of a
real number $x$.

\section{System Model} \label{sec:system model}

\begin{figure}
\psfrag{a}{Base
Station}\psfrag{b}{User-1}\psfrag{c}{User-2}\psfrag{d}{User-K}
\begin{center}
\scalebox{1.0}{\includegraphics*[36pt,533pt][287pt,728pt]{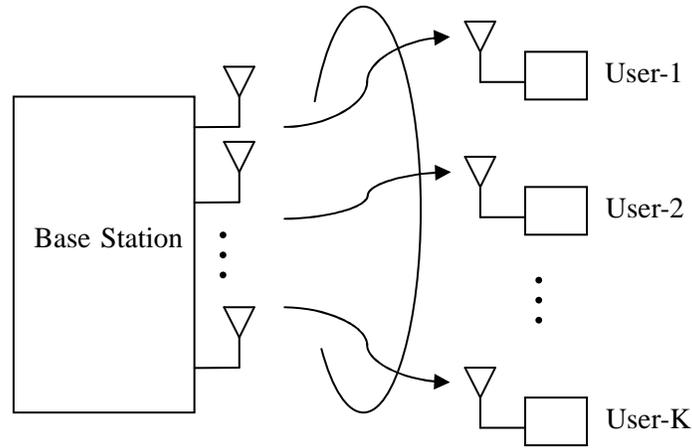}}
\caption{MISO-BC for SDMA-based downlink transmission in a
single-cell of wireless cellular network. }\label{fig:system model}
\end{center}
\end{figure}

A MISO-BC channel with $K$ mobile users each having a single antenna
and a fixed BS having $M$ antennas is considered, as shown in Fig.
\ref{fig:system model}. Because of multi-antennas at the
transmitter, the BS is able to employ SDMA to transmit to multiple
users simultaneously at the same bandwidth. It is assumed that the
transmission to all users is synchronously divided into consecutive
blocks, and the fading occurs from block to block but remains static
within a block of symbols, i.e., a block-fading (BF) model.
Furthermore, it is assumed that the fading process is stationary and
ergodic. Let $n$ be the random variable representing the fading
state. At fading state $n$, the MISO-BC can be considered as a
discrete-time channel represented by
\begin{eqnarray} \label{eq:MIMO BC}
\left[\begin{array}{c} y_1(n)\\ \vdots \\
y_K(n)\end{array} \right]=\left[\begin{array}{c} \mv{h}_1(n)
\\ \vdots
\\ \mv{h}_K(n)\end{array}\right] \mv{x}(n)+\left[
\begin{array}{c} z_1(n) \\ \vdots \\ z_K(n) \end{array}
\right],
\end{eqnarray}
where $y_k(n)$, $\mv{h}_k(n)$ and $z_k(n)$ denote the received
signal, the $1\times M$ downlink channel vector, and the receiver
noise for user $k$, respectively, and
$\mv{x}(n)\in\mathbb{C}^{M\times 1}$ denotes the transmitted signal
vector from the BS. It is assumed that $z_k(n)\sim\mathcal{CN}(0,1),
\forall n, k$. The transmitted signal $\mv{x}(n)$ can be further
expressed as
\begin{eqnarray} \label{eq:x(n)}
\mv{x}(n)=\sum_{k=1}^{K}\mv{b}_k(n)s_k(n),
\end{eqnarray}
where $\mv{b}_k(n)\in\mathbb{C}^{M\times 1}$ and $s_k(n)$ represent
the precoding vector and the transmitted codeword symbol for user
$k$, respectively, at fading state $n$. It is assumed that each user
employs the optimal Gaussian code-book with normalized codeword
symbols, i.e., $s_k(n)\sim\mathcal{CN}(0,1), \forall k,n$, and the
rate of code-book for user $k$ at fading state $n$ is denoted as
$r_k(n)$. The allocated transmit power to user $k$ at fading state
$n$ is denoted by $p_k(n)$, and it can be easily verified that
$p_k(n)=\|\mv{b}_k(n)\|^2$. The total transmit power from the BS at
fading state $n$ is then expressed as $p(n)=\sum_{k=1}^{K}p_k(n)$.
Assuming full knowledge of the fading distribution, the BS is able
to adapt the transmission power $p_k(n)$ and rate $r_k(n)$ (could be
both zero for some fading state $n$) allocated to user $k$ in order
to exploit multiuser channel variations over time. Assuming a
long-term power constraint (LTPC) $p^*$ over different fading
states, the average transmit power at the BS needs to satisfy
$\mathbb{E}_n[p(n)]\leq p^{*}$.

Supposing that $p(n)$ is given, the achievable rates $\{r_k(n)\}$ of
users need to be contained in the corresponding capacity region of
the MISO-BC at fading state $n$, denoted by $\mathcal{C}^{\rm
BC}_n(p(n),\{\mv{h}_k(n)\})$. Characterization of $\mathcal{C}^{\rm
BC}_n$ will become useful later in this paper when the issue on how
to dynamically allocate transmit power and user rates at different
fading states is addressed. In many cases, it is more convenient to
apply the celebrated duality result between the Gaussian BC and MAC
\cite{Goldsmith03} to transform the capacity region characterization
for the original BC to that for its dual MAC. Assuming that in the
dual SIMO-MAC of the original MISO-BC considered in this paper, each
user employs the optimal Gaussian code-book of rate $R_k(n)$ and has
a transmit power $q_k(n)$, $k=1,\ldots,K$, at fading state $n$, by
\cite{Coverbook} the capacity region of the dual SIMO-MAC at fading
state $n$ can be expressed as
\begin{eqnarray}\label{eq:capacity region dual MAC}
\mathcal{C}^{\rm MAC}_n(\{q_k(n)\},\{\mv{h}_k^{\dag}(n)\})= \left\{
\mv{R}(n)\in \mathbb{R}_{+}^K: \sum_{k\in J}R_k(n) \leq \log_2
\left|\sum_{k\in J}\mv{h}_k^{\dag}(n)\mv{h}_k(n)
q_k(n)+\mv{I}\right|, \forall J\subseteq \{1,\ldots,K\}\right\},
\end{eqnarray}
where $\mv{R}(n)\triangleq[R_1(n),\ldots,R_K(n)]$. The duality
result \cite{Goldsmith03} then states that an achievable rate region
for the original MISO-BC at fading state $n$ with total transmit
power $p(n)$ can be expressed as

\begin{equation}\label{eq:Rate MIMO BC}
\mathcal{R}^{\rm BC}_n (p(n), \{\mv{h}_k(n)\}) =
\bigcup_{\{q_k(n)\}: q_1(n)+\ldots + q_K(n) \leq p(n)}
\mathcal{C}^{\rm MAC}_n(\{q_k(n)\},\{\mv{h}_k^{\dag}(n)\}).
\end{equation}

It was later shown in \cite{Shamai04} that the above achievable rate
region $\mathcal{R}^{\rm BC}_n(p(n), \{\mv{h}_k(n)\})$ is indeed the
capacity region $\mathcal{C}^{\rm BC}_n(p(n), \{\mv{h}_k(n)\})$ for
the Gaussian BC. By applying the above results, it follows that any
rate-tuple $\{r_k(n)\}$ that is achievable in the fading MISO-BC by
the user power-tuple $\{p_k(n)\}$ is also achievable as
$\{R_k(n)\}$, $R_k(n)=r_k(n), \forall k, n$, in the dual fading
SIMO-MAC by the corresponding user power-tuple $\{q_k(n)\}$ provided
that $\sum_{k=1}^{K}p_k(n)=\sum_{k=1}^{K}q_k(n), \forall n$. Note
that the power allocation $p_k(n)$ for user $k$ in the original BC
is not necessarily equal to $q_k(n)$ in the dual MAC. The transforms
between $\{q_k(n)\}$ and $\{p_k(n)\}$ as well as the corresponding
precoding vectors $\{\mv{b}_k(n)\}$ for the same set of achievable
rates $\{r_k(n)\}$ and $\{R_k(n)\}$ can be found in
\cite{Goldsmith03}, and are thus omitted in this paper for brevity.

\section{Dynamic Resource Allocation under Heterogeneous Delay Constraints} \label{sec:mixed traffic}

This section studies optimal dynamic resource allocation algorithms
for the BF MISO-BC to support simultaneous transmission of data
traffic with heterogeneous transmit rate and delay constraints.
First, Section \ref{subsec:problem formulation} provides the problem
formulation. Then, Section \ref{subsec:proposed solution} presents
the solution based on the Lagrange-duality method of convex
optimization. At last, Section \ref{subsec:implementation} derives
an online algorithm that is suitable for real-time implementation of
the proposed solution.

\subsection{Problem Formulation} \label{subsec:problem formulation}

\begin{figure}
\psfrag{a}{Packet Data}\psfrag{b}{Variable Transmit Rate
}\psfrag{c}{Circuit Data}\psfrag{d}{Constant Transmit
Rate}\psfrag{e}{Higher Layer Application}\psfrag{f}{Physical-Layer
Transmission}
\begin{center}
\scalebox{1.0}{\includegraphics*[57pt,565pt][483pt,739pt]{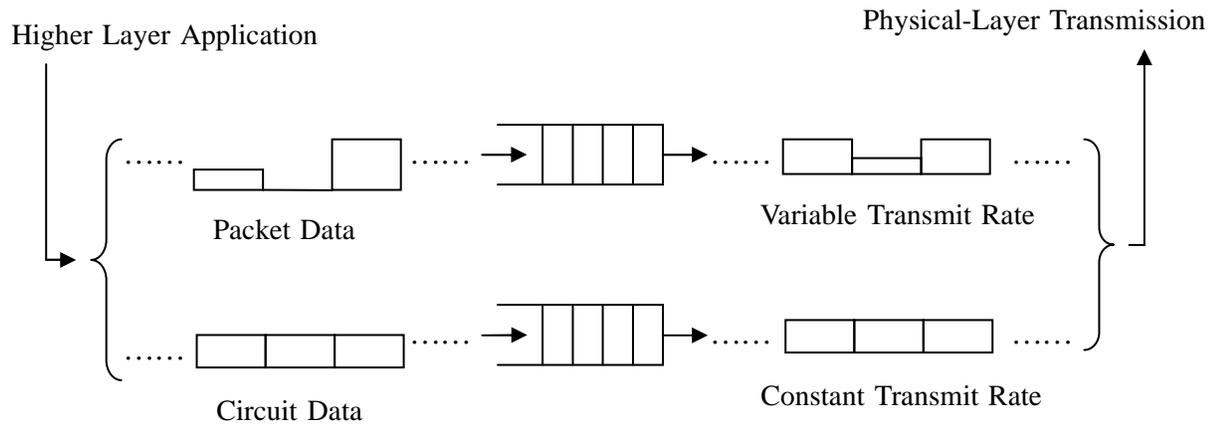}}
\caption{Transmission scheduling at the MAC-layer for packet data
and circuit data.}\label{fig:traffic model}
\end{center}
\end{figure}

The following rule for transmission scheduling at the BS is
considered. As illustrated in Fig. \ref{fig:traffic model}, each
user's data arising from some higher layer application is first
placed into a dedicated buffer. Periodically, the BS removes some of
the data from each user's buffer, jointly encodes them into a block
of symbols, and then broadcasts the encoded block to all users
through the MISO-BC. For simplicity, it is assumed that all user's
data arrive to their dedicated buffers synchronously at the
beginning of each scheduling period. The data arrival processes of
users are assumed to be stationary and ergodic, mutually
independent, and also independent of their channel realizations.
This paper considers two types of data traffic with very different
delay requirements: One is the delay-tolerant packet data and the
other is the delay-sensitive circuit data, for which the following
assumptions are made:
\begin{itemize}
\item For a user with packet data application, the data arrival process
is not necessarily continuous in time and the amount of arrived data
in each scheduling period may be variable. All data are stored in a
buffer of a sufficiently large size such that data dropping due to
buffer overflow does not occur. In order for the scheduler to
optimally exploit the channel dynamics, the allocated transmit rate
can be variable during each scheduling period. It is assumed that
there is always a sufficient amount of backlogged data in the buffer
for transmission. The scheduler needs to ensure that the transmit
rate averaged over scheduling periods in the long run is no smaller
than the average data arrival rate. However, the exact amount of
delay incurred to transmitted data in the buffer is not guaranteed.

\item For a user with circuit data application, the data arrival
process is continuous with a constant-rate during each scheduling
period. The arrived data is stored in the buffer for only one
scheduling period and then transmitted. Therefore, the amount of
delay incurred to transmitted data is minimal. However, the
scheduler needs to ensure a constant-rate transmission independent
of channel condition.
\end{itemize}

Let the users with packet data applications be represented by the
set $\mathcal{U}_{\rm NDC}$ where NDC refers to
no-delay-constrained, and the users with circuit data applications
represented by $\mathcal{U}_{\rm DC}$ where DC refers to
delay-constrained. In this paper, we consider optimal dynamic
resource allocation to minimize the average transmit power at the BS
over different fading states subject to the constraint that all NDC
and DC user rate demands are satisfied. Recall that $r_k(n)$ denotes
the rate assigned to user $k$ by the scheduler at fading state $n$.
For a NDC user, it is required that the average transmit rate
$\mathbb{E}_n[r_k(n)]$ over fading states needs to be no smaller
than its average data arrival rate $R_k^*$. In contrast, for a DC
user, the transmit rate $r_k(n)$ at any fading state $n$ needs to
satisfy its constant data arrival rate $R_k^*$. By considering the
dual SIMO-MAC in Section \ref{sec:system model} with $R_k(n)=r_k(n),
q_k(n)=p_k(n), \forall k, n$, optimal allocation of transmit rates
and powers at different $n$ can be obtained by solving the following
optimization problem ({\bf P1}):
\begin{eqnarray}
\mathtt{Minimize} && \mathbb{E}_n\left[\sum_{k=1}^K q_k(n)\right] \\
\mathtt {Subject \ to} && \mathbb{E}_n[R_k(n)] \geq R_k^{*}, \ \forall k\in \mathcal{U}_{\rm NDC} \label{eq:VR constraint} \\
&& R_k(n) \geq R_k^{*}, \ \forall n, \ \forall k\in \mathcal{U}_{\rm DC} \label{eq:CR constraint} \\
&& \mv{R}(n)\in\mathcal{C}^{\rm MAC}_n\left(\{q_k(n)\},
\{\mv{h}_k^{\dag}(n)\}\right), \ \forall n \label{eq:capacity
constraint}
\\&& q_k(n)\geq 0 \ \forall n,k. \label{eq:power constraint}
\end{eqnarray}
For the above problem, the objective function and all the
constraints except (\ref{eq:capacity constraint}) are affine. It can
also be verified from (\ref{eq:capacity region dual MAC}) that
$\mathcal{C}^{\rm MAC}_n$ is a convex set with any given positive
$\{q_k(n)\}$. Therefore, Problem P1 is a convex optimization problem
\cite{Boydbook} and thus can be solved using efficient convex
optimization techniques, as will be shown next.

\subsection{Proposed Solution} \label{subsec:proposed solution}
The Lagrange-duality method is usually applied when a convex
optimization problem can be more conveniently solved in its dual
domain than in its original form. In this paper, we also apply this
method for solving Problem P1. The first step for the
Lagrange-duality method is to introduce dual variables associated
with some constraints of the original problem. For Problem P1 that
has multiple constraints, there are also various ways to introduce
dual variables that might result in different dual problems. For the
following proposed solution, dual variables are chosen with an aim
to facilitate implementing it in the real time, as will be explained
later in Section \ref{subsec:implementation}.

As a first step, a set of dual variables $\{\mu_k\}$, $\mu_k\geq 0,
k\in\mathcal{U}_{\rm NDC}$, are introduced for the NDC users with
respect to (w.r.t.) their average-rate constraints in (\ref{eq:VR
constraint}). The Lagrangian of Problem P1 can be then expressed as
\begin{eqnarray}\label{eq:Lagrange power min}
\mathcal{L}(\{q_k(n)\},\{R_k(n)\},\{\mu_k\})
=\mathbb{E}_n\left[\sum_{k=1}^K
q_k(n)\right]-\sum_{k\in\mathcal{U}_{\rm
NDC}}\mu_k\left(\mathbb{E}_n[R_k(n)]- R_k^{*}\right).
\end{eqnarray}
Denote the set of $\{q_k(n)\}$ and $\{R_k(n)\}$ specified by the
remaining constraints in (\ref{eq:CR constraint}), (\ref{eq:capacity
constraint}) and (\ref{eq:power constraint}) as $\mathcal{D}$, the
Lagrange dual function is expressed as
\begin{eqnarray}\label{eq:Lagrange dual power min}
g(\{\mu_k\})=\min_{\left\{q_k(n),R_k(n)\right\} \in \mathcal{D}}
\mathcal{L}(\{q_k(n)\},\{R_k(n)\},\{\mu_k\}).
\end{eqnarray}
The dual problem of the original (primal) problem can be then
expressed as
\begin{equation} \label{eq:dual problem}
\max_{\mu_k\geq 0, k\in\mathcal{U}_{\rm NDC}} g(\{\mu_k\}).
\end{equation}
Because the primal problem is convex and also satisfies the Slater's
condition \cite{Boydbook},\footnote{Slater's condition requires that
the feasible set of the optimization problem has non-empty interior,
which is in general the case for Problem P1 because with
sufficiently large average transmit power, any finite user rates
$\{R_k^*\}$ are achievable.} the duality gap between the optimal
value of the primal problem and that of the dual problem becomes
zero. This suggests that the problem at hand can be equivalently
solved in its dual domain by first minimizing the Lagrangian
$\mathcal{L}$ to obtain the dual function $g(\{\mu_k\})$ for some
given $\{\mu_k\}$, and then maximizing $g(\{\mu_k\})$ over
$\{\mu_k\}$.

Considering first the minimization problem in (\ref{eq:Lagrange dual
power min}) to obtain $g(\{\mu_k\})$ for some given $\{\mu_k\}$. It
is interesting to observe that this problem can be solved by
considering parallel subproblems each corresponding to one fading
state $n$. From (\ref{eq:Lagrange power min}), the subproblem for
fading state $n$ can be written as ({\bf P2})
\begin{eqnarray}
\mathtt{Minimize} && \sum_{k=1}^K q_k(n)- \sum_{k\in\mathcal{U}_{\rm
NDC}}\mu_kR_k(n) \\
\mathtt {Subject \ to}
&& R_k(n) \geq R_k^{*}, \ \forall k\in \mathcal{U}_{\rm DC} \label{eq:CR constraint sub} \\
&& \mv{R}(n)\in\mathcal{C}^{\rm MAC}_n\left(\{q_k(n)\},
\{\mv{h}_k^{\dag}(n)\}\right) \label{eq:capacity constraint sub}
\\&& q_k(n)\geq 0 \ \forall k. \label{eq:power constraint sub}
\end{eqnarray}
Hence, the dual function $g(\{\mu_k\})$ can be obtained by solving
subproblems all having the identical structure, a technique usually
referred to as the {\it Lagrange-dual decomposition}. For solving
Problem P2 for each $n$, a new set of positive dual variables
$\delta_k(n), k\in\mathcal{U}_{\rm DC}$, are introduced for DC users
w.r.t. their constant-rate constraints in (\ref{eq:CR constraint
sub}). The Lagrangian of Problem P2 can be then expressed as
\begin{eqnarray}\label{eq:Lagrange power min sub}
\mathcal{L}_n(\{q_k\},\{R_k\},\{\delta_k)\}) =\sum_{k=1}^K
q_k-\sum_{k\in\mathcal{U}_{\rm
NDC}}\mu_kR_k-\sum_{k\in\mathcal{U}_{\rm DC}}\delta_k(R_k-R_k^*).
\end{eqnarray}
Note that for brevity, the index $n$ is dropped in $q_k(n)$,
$R_k(n)$ and $\delta_k(n)$ in (\ref{eq:Lagrange power min sub})
since it is applicable for all $n$. The corresponding dual function
can be then defined as
\begin{eqnarray}\label{eq:Lagrange dual power min sub}
g_n(\{\delta_k\})=\min_{\left\{q_k,R_k\right\} \in \mathcal{D}_n}
\mathcal{L}_n(\{q_k\},\{R_k\},\{\delta_k\}),
\end{eqnarray}
where $\mathcal{D}_n$ denotes the set of $\{q_k\}$ and $\{R_k\}$
specified by the remaining constraints (\ref{eq:capacity constraint
sub}) and (\ref{eq:power constraint sub}) at fading state $n$. The
associated dual problem is then defined as
\begin{equation} \label{eq:dual problem sub}
\max_{\delta_k\geq 0, k\in\mathcal{U}_{\rm DC}} g_n(\{\delta_k\}).
\end{equation}
Similar like P1, Problem P2 can also be solved by first minimizing
$\mathcal{L}_n$ to obtain the dual function $g_n(\{\delta_k\})$ for
a given set of $\{\delta_k\}$, and then maximizing $g_n$ over
$\{\delta_k\}$. From (\ref{eq:Lagrange power min sub}), the
minimization problem in (\ref{eq:Lagrange dual power min sub}) can
be expressed as
\begin{eqnarray}
\mathtt{Minimize} && \sum_{k=1}^K q_k- \sum_{k\in\mathcal{U}_{\rm
NDC}}\mu_kR_k-\sum_{k\in\mathcal{U}_{\rm
DC}}\delta_kR_k\\
\mathtt {Subject \ to} && \mv{R}\in\mathcal{C}^{\rm
MAC}_n\left(\{q_k\}, \{\mv{h}_k^{\dag}\}\right) \label{eq:capacity
constraint sub sub}
\\&& q_k\geq 0 \ \forall k. \label{eq:power constraint sub sub}
\end{eqnarray}
Let $\beta_k=\mu_k$, if $k\in \mathcal{U}_{\rm NDC}$, and
$\beta_k=\delta_k$, if $k\in \mathcal{U}_{\rm DC}$, and $\pi$ be a
permutation over $\{1,\ldots,K\}$ such that $\beta_{\pi(1)}\geq
\beta_{\pi(2)} \geq \cdots \geq \beta_{\pi(K)}$, and let
$\beta_{\pi(K+1)}\triangleq 0$. Thanks to the polymatroid structure
of $\mathcal{C}^{\rm MAC}_n$ \cite{Tse98a}, the above problem can be
simplified as ({\bf P3})
\begin{eqnarray}
\mathtt{Minimize} && \sum_{k=1}^K q_k- \sum_{k=1}^{K}
\left(\beta_{\pi(k)}-\beta_{\pi(k+1)}\right)\log_2\left|\sum_{i=1}^{k}\mv{h}_{\pi(i)}^{\dag}\mv{h}_{\pi(i)}q_{\pi(i)}
+\mv{I}\right| \label{eq:objective func}\\ \mathtt {Subject \ to} &&
q_k\geq 0 \ \forall k. \label{eq:power constraint sub sub}
\end{eqnarray}
Problem P3 is convex because all the constraints are affine and the
objective function is convex w.r.t. $\{q_k\}$. Hence, this problem
can be solved, e.g., by the interior-point method \cite{Boydbook}.
In Appendix \ref{appendix:algorithm}, an alternative method based on
the block-coordinate decent principle \cite{Bertsekasbook} by
iteratively optimizing $q_k$ with all the other $\{q_{k'}\}, k'\neq
k$ as fixed is presented. This method can be considered as a
generalization of the algorithm described in \cite{Yu03}, where all
$\beta_{\pi(k)}, k=1,\ldots,K$, are equal.

So far, solutions have been presented for the minimization problem
in (\ref{eq:Lagrange dual power min}) to obtain $g(\{\mu_k\})$ for
some given $\{\mu_k\}$ and that in (\ref{eq:Lagrange dual power min
sub}) to obtain each $g_n(\{\delta_k\})$ for some given
$\{\delta_k\}$. Next, the remaining issue on how to update
$\{\mu_k\}$  to maximize $g(\{\mu_k\})$ for the dual problem in
(\ref{eq:dual problem}) is addressed. Similar techniques can also be
used for updating $\{\delta_k\}$ to maximize $g_n(\{\delta_k\})$ for
each dual problem in (\ref{eq:dual problem sub}). From
(\ref{eq:Lagrange power min}) and (\ref{eq:Lagrange dual power
min}), it is observed that the dual function $g(\{\mu_k\})$, though
affine w.r.t. $\{\mu_k\}$, is not directly differentiable w.r.t.
$\{\mu_k\}$. Hence, standard method like Newton method cannot be
employed to update $\{\mu_k\}$ for maximizing $g(\{\mu_k\})$. An
appropriate choice here may be the {\it sub-gradient-based} method
\cite{Bertsekasbook} that is capable of handling non-differentiable
functions. This method is an iterative algorithm and at each
iteration, it requires a sub-gradient at the corresponding value of
$\{\mu_k\}$ to update $\{\mu_k\}$ for the next iteration. Suppose
that after solving Problem P2 for some given $\{\mu_k\}$ at all
fading states of $n$, the obtained rates and powers are denoted by
$\{R'_k(n)\}$ and $\{q'_k(n)\}$, respectively, $k\in\mathcal{U}_{\rm
NDC}$. The following lemma then provides a suitable sub-gradient for
$\{\mu_k\}$:
\begin{lemma}\label{lemma:sub-gradient}
If $\mathcal{L}(\{q'_k(n)\},\{R'_k(n)\},\{\mu_k\})= g(\{\mu_k\})$,
then the vector $\mv{\nu}$ defined as $\nu_k = R_k^* -
\mathbb{E}_n[R'_k(n)]$ for $k\in\mathcal{U}_{\rm NDC}$ is a
sub-gradient of $g$ at $\{\mu_k\}$.
\end{lemma}
\begin{proof}
Since for any set of $\theta_k$'s, $\theta_k\geq 0,
k\in\mathcal{U}_{\rm NDC}$, it follows that
\begin{eqnarray}
g(\{\theta_k\}) &\leq&
\mathcal{L}(\{q'_k(n)\},\{R'_k(n)\},\{\theta_k\})\\ &=& g(\{\mu_k\})
+ \sum_{k\in\mathcal{U}_{\rm NDC}} (\theta_k - \mu_k)\left( R_k^* -
\mathbb{E}_n[R'_k(n)]\right).
\end{eqnarray}
Hence, it is clear that the optimal dual solution $\{\mu_k^*\}$ that
maximizes $g(\{\mu_k\})$ should not lie in the half space
represented by $\{\mv{\theta}: \sum_{k\in\mathcal{U}_{\rm NDC}}
(\theta_k - \mu_k)\nu_k\leq 0\}$. As a result, any new update of
$\{\mu_k\}$ for the next iteration, denoted by $\{\mu^{\rm
new}_k\}$, should satisfy $\sum_{k\in\mathcal{U}_{\rm NDC}}
(\mu^{\rm new}_k - \mu_k)\nu_k\geq 0$.
\end{proof}
By applying Lemma \ref{lemma:sub-gradient}, a simple rule for
updating $\{\mu_k\}$ is as follows:\footnote{Notice that a more
efficient sub-gradient-based method to iteratively find
$\{\mu_k^*\}$ is the ellipsoid method \cite{BGT81}, which at each
iteration removes the half space specified by $\{\mv{\theta}:
\sum_{k\in\mathcal{U}_{\rm NDC}} (\theta_k - \mu_k)\nu_k\leq 0\}$
for searching $\{\mu_k^*\}$.}
\begin{eqnarray} \label{eq:mu update}
\mu^{\rm new}_k=\left\{\mu_k+\Delta\left(R_k^* -
\mathbb{E}_n[R'_k(n)]\right)\right\}^+, \ \ k\in\mathcal{U}_{\rm
NDC},
\end{eqnarray}
where $\Delta$ is a small positive step size. Similarly, at each
fading state $n$, $\{\delta_k(n)\}$ can be also updated towards
maximizing $g_n(\{\delta_k(n)\})$ as
\begin{eqnarray} \label{eq:delta update}
\delta^{\rm new}_k(n)=\left\{\delta_k(n)+\Delta\left(R_k^* -
R'_k(n)\right)\right\}^+, \ \ k\in\mathcal{U}_{\rm DC},
\end{eqnarray}
where $\{R'_k(n)\}$, $k\in\mathcal{U}_{\rm DC}$, is the solution
obtained after solving Problem P3 at fading state $n$.

To summarize, the proposed solution is implemented by a {\it
two-layer} Lagrange-duality method. At the first layer, the
algorithm searches iteratively for $\{\mu_k^*\}$ with which the
average-rate constraints of all NDC users are satisfied. At each
iteration, an update for $\{\mu_k\}$ is generated and then passed to
the second layer where the algorithm starts a parallel search for
$\{\delta_k^*(n)\}$, each for a fading state $n$, such that the
constant-rate constraints of all DC users are satisfied at all
fading states. The resultant $\{R'_k(n)\}$ of NDC users is then
passed back to the first layer for another update of $\{\mu_k\}$.
The overall algorithm is summarized in Table \ref{table:minpower
algorithm}. The complexity of this algorithm can be derived as
follows. Supposing that the ellipsoid method is used to iteratively
update dual variables, the required number of iterations for
convergence has $\mathcal{O}(m^2)$, where $m$ is the size of the
problem. Let $K_{\rm NDC}$ and $K_{\rm DC}$ denote the size of
$\mathcal{U}_{\rm NDC}$ and $\mathcal{U}_{\rm DC}$, respectively,
where $K_{\rm NDC}+K_{\rm DC}=K$. Therefore, the ellipsoid method
will need $\mathcal{O}(K_{\rm NDC}^2)$ iterations for obtaining
$\{\mu_k^*\}$ and $\mathcal{O}(NK_{\rm DC}^2)$ iterations for
obtaining $\{\delta_k^*(n)\}$ for all fading states, assuming the
number of fading states $n$ is finite and is equal to $N$.  The
complexity for solving Problem P3 is $\mathcal{O}(K^2)$ by, e.g.,
the interior-point method. Hence, the total complexity of the
algorithm is $\mathcal{O}(K_{\rm NDC}^2K_{\rm DC}^2K^2N)$.

At last, take note that the proposed algorithm jointly optimizes
transmit powers and rates together with decoding orders (determined
by the magnitudes of $\{\mu^*_k\}$ and $\{\delta_k^*(n)\}$
\cite{Tse98a}) of users at all fading states for the dual fading
SIMO-MAC. By the BC-MAC duality result \cite{Goldsmith03}, the
optimal transmit powers, rates, precoding vectors as well as
encoding orders of users for the original fading MISO-BC can be
obtained.

\subsection{Online Algorithm}\label{subsec:implementation}

One important issue yet to be addressed for implementing the
proposed solution for Problem P1 is how to relax its assumption on
perfect knowledge of the distribution of fading state $n$. Notice
that this knowledge is necessary for computing the achievable
average rates $\{\mathbb{E}_n[R'_k(n)]\}$, $k\in\mathcal{U}_{\rm
NDC}$, which are needed for updating the dual variables $\{\mu_k\}$
for NDC users in (\ref{eq:mu update}). In practice, although it is
reasonable to assume each user's fading channel is stationary and
ergodic, the space of fading states is usually continuous and
infinite and hence it is infeasible for the BS to initially acquire
the channel distribution information for all users at all fading
states. Even though this information is available for off-line
implementation of the proposed solution, the computational
complexity becomes unbounded as the number of fading states goes to
infinity. Therefore, in this paper a modified ``online'' algorithm
is developed that is able to adaptively update $\{\mu_k\}$ towards
$\{\mu_k^*\}$ as transmission proceeds over time. Let $t$ denote the
transmission block index, $t=1,2,\cdots$. The key for the online
algorithm is to approximate the statistical average
$\mathbb{E}_n[R'_k(n)]$, $k\in\mathcal{U}_{\rm NDC}$, at time $t$ by
a time average of transmitted rates up to time $t-1$, denoted by
$\bar{R}_k[t-1]$, where $\bar{R}_k[t]$ is obtained as
\begin{equation}\label{eq:rate update}
\bar{R}_k[t]=(1-\epsilon)\bar{R}_k[t-1]+\epsilon R_k[t],
\end{equation}
where $R_k[t]$ is the transmitted rate at time $t$, and $\epsilon$,
$0<\epsilon\ll1$, is a parameter that controls the convergence speed
of $\bar{R}_k[t]\rightarrow \mathbb{E}_n[R'_k(n)]$ as $t\rightarrow
\infty$. By replacing $\mathbb{E}_n[R'_k(n)]$ by $\bar{R}_k[t-1]$ in
(\ref{eq:mu update}), $\{\mu_k[t]\}$ at time $t$ can be updated
accordingly such that as $t\rightarrow \infty$ it converges to
$\{\mu_k^*\}$ because $\bar{R}_k[t]\rightarrow R_k^*, \forall
k\in\mathcal{U}_{\rm NDC}$. This modified online algorithm is
summarized in Table \ref{table:online algorithm}.

{\bf Cross-Layer Implementation:} The proposed online algorithm
based on the Lagrange-duality method is amenable to cross-layer
implementation of both PHY-layer transmission and MAC-layer
multiuser rate scheduling. One challenging issue for cross-layer
optimization is on how to select useful and succinct information for
different layers to exchange and share so as to optimize their
individual operations. The Lagrange-duality method provides a new
design paradigm for efficient cross-layer information exchange. On
the one hand, since the MAC-layer has the knowledge of user rate
demands $\{R_k^*\}$ as well as their delay requirements (NDC or DC),
it can update dual variables $\{\mu_k[t]\}$ and $\{\delta_k[t]\}$
accordingly (see Table \ref{table:online algorithm}), and then pass
them to the PHY-layer for computing the desirable transmission rates
of users $\{R_k[t]\}$ to meet with each user's specific rate demand.
On the other hand, the PHY-layer is able to provide the MAC-layer
the updated $\{R_k[t]\}$ that optimize the PHY-layer transmission by
solving Problem P3 given $\{\mu_k[t]\}$ and $\{\delta_k[t]\}$.
Therefore, by exchanging dual variables and transmission rates
between PHY- and MAC-layers, cross-layer dynamic resource allocation
can be efficiently implemented.

{\bf Comparison with PFS:} It is interesting to draw a comparison
between the proposed online algorithm and the well-known PFS
algorithm. PFS is designed for real-time multiuser rate scheduling
in a mobile wireless network to ensure some certain fairness for
user rate allocation while maximizing the network throughput. PFS
applies to packet data transmission and hence is equivalent to the
transmission scenario considered in this paper when only NDC users
are present. At each time $t$, transmit rates $\{R_k[t]\}$ assigned
to users by PFS maximize the weighted sum-rate of users
$\sum_k\omega_k[t]R_k[t]$, where the weights are given by
$\omega_k[t]=\frac{1}{\bar{R}_k[t-1]}$ and $\bar{R}_k[t-1]$ is the
estimated average rate for user $k$ up to time $t-1$ the same as
expressed in (\ref{eq:rate update}). Using this rule, it has been
shown (e.g., \cite{Tse02}-\cite{Caire07} and references therein)
that as $t\rightarrow \infty$, $\bar{R}_k[t]\rightarrow R_k^*$ where
$R_k^*$ is the average achievable rate for user $k$ in the long
term. Furthermore, PFS maximizes $\sum_k\log(R_k^*)$ over the
expected capacity region (please refer to Definition
\ref{def:capacity region} in Section \ref{sec:tradeoffs}) and,
hence, $\{R_k^*\}$ can be considered as the unique intersection of
the surface specified by $\prod_{k}R_k=c$ and the boundary of the
expected capacity region. Because of the $\log(\cdot)$ function, the
rate assignments among users by PFS are regulated in a balanced
manner such that no user can be allocated an overwhelmingly larger
rate than the others even if it has a superior channel condition.
However, the achievable rates $\{R_k^*\}$ by PFS are not guaranteed
to satisfy any desired rate demand of users. In contrast, the
proposed online algorithm ensures that each NDC user's average-rate
demand is satisfied by applying a different rule (see Table
\ref{table:online algorithm}) for updating user weights
$\{\mu_k[t]\}$ for the resource allocation problem (see Problem P3)
to be solved at each $t$.

\section{Throughput-Delay Tradeoff}
\label{sec:tradeoffs}

For a single-user BF channel, the expected capacity
\cite{Goldsmith97}, \cite{Berry02} and the delay-limited capacity
\cite{Tse98b} can be considered as the fading channel capacity
limits under two extreme cases of delay constraint. The former is
always larger than the latter and their difference, termed the {\it
delay penalty}, then characterizes a fundamental throughput-delay
tradeoff for dynamic resource allocation over a single-user fading
channel. The delay penalty may or may not be significantly large for
a single-user fading channel. For example, for a SISO-BF channel,
the delay penalty can be substantial because the delay-limited
capacity is indeed zero if the fading channel is not ``invertible''
with a finite average transmit power \cite{Caire99}. However, when
multi-antennas are employed at the transmitter and/or receiver, the
delay-limited capacity of the MIMO-BF channel can be very close to
the channel expected capacity \cite{Caire00}, i.e., a negligible
delay penalty. This result can be explained by the
``channel-hardening'' effect \cite{Tarokh04} for random MIMO
channels, i.e., the mutual information of independent MIMO channels
becomes less variant because of antenna-induced space diversity, and
hence the value of power and rate adaptation over time vanishes as
the number of antennas becomes large. This result indicates that
from an information-theoretic viewpoint, MIMO channels are highly
suitable for transmission of real-time and delay-constrained data
traffic.

Characterization of the delay penalty in a multiuser fading channel
is more challenging. The capacity definitions for the single-user
fading channel can be extended to the multiuser channel as the {\it
expected capacity region} under no delay constraint for all users,
and the {\it delay-limited capacity region} under zero-delay
constraint for all users. Therefore, the delay penalty can be
measured by directly comparing these two capacity regions. Since
capacity region contains the achievable rates of all users, it lies
in a $K$-dimensional space where $K$ is the number of users in the
network. As a result, characterization of capacity region becomes
inconvenient as $K$ becomes large. In order to overcome this
difficulty, prior research work usually adopts the maximum sum-rate
of users over the capacity region, termed the {\it sum capacity}, as
a simplified measure for the network throughput. Applying the sum
capacity to the corresponding capacity region, the {\it expected
throughput} and the {\it delay-limited throughput} can be defined
accordingly. However, the conventional sum capacity does not
consider the rate allocation among users. As a consequence, each
user's allocated rate portion in the expected throughput may be
different compared to that in the delay-limited throughput. Hence, a
similar measure for the delay penalty like in the single-user case
by taking the difference between the expected and delay-limited
throughput looks problematic at a first glance in the multiuser
case.

This section presents a novel characterization of the fundamental
throughput-delay tradeoff for the fading MISO-BC. Instead of
considering mixed NDC and DC transmission like Section
\ref{sec:mixed traffic}, it is assumed here that there are only NDC
or DC users present, and comparison of the network throughput under
these two extreme cases of delay constraint is of interest. Because
it is hard to compare directly the expected and the delay-limited
capacity region as the number of users becomes large, the sum
capacity is also considered for simplicity. However, unlike the
conventional sum capacity that does not guarantee the amount of rate
allocation among users, the expected and delay-limited throughput in
this paper are defined under a new constraint that regulates each
user's allocated rate portion based upon a prescribed {\it
rate-profile}. Then, the delay penalty is characterized by the
difference between the expected and delay-limited throughput {\it
under the same rate-profile}. The above concepts are more explicitly
defined as follows:
\begin{definition}\label{def:capacity region}
The expected capacity region for a fading MISO-BC expressed in
(\ref{eq:MIMO BC}) under a LTPC $p^*$ can be defined as
\begin{eqnarray}\label{eq:expected capacity region}
\mathcal{C}_{\rm e}(p^*) = \left \{\mv{r}\in\mathbb{R}^K_+:
r_k=\mathbb{E}_n[R_k(n)], \forall k, \mv{R}(n)\in\mathcal{C}^{\rm
BC}_n(p(n),\{\mv{h}_k(n)\}), \forall n, \mathbb{E}_n[p(n)]\leq
p^*\right\}.
\end{eqnarray}
Similarly, the delay-limited capacity region is defined as
\begin{eqnarray}\label{eq:DLC region}
\mathcal{C}_{\rm d}(p^*) = \left \{\mv{r}\in\mathbb{R}^K_+:
r_k=R_k(n), \forall k,n, \mv{R}(n)\in\mathcal{C}^{\rm
BC}_n(p(n),\{\mv{h}_k(n)\}), \forall n, \mathbb{E}_n[p(n)]\leq
p^*\right\}.
\end{eqnarray}
\end{definition}

\begin{definition}
Let $R_k^*$ denote $k$-th user's rate demand (average-rate for a NDC
user or constant-rate for a DC user) $k=1,\ldots,K$, the rate
profile is defined as a vector
$\mv{\alpha}=\{\alpha_1,\ldots,\alpha_K\}$, where
$\alpha_k=\frac{R_k^*}{\sum_{k=1}^{K}R_k^*}, k=1,\ldots,K$.
\end{definition}

\begin{definition}\label{def:throughput}
The expected throughput $C_{\rm e}(p^*,\mv{\alpha})$ (delay-limited
throughput $C_{\rm d}(p^*,\mv{\alpha})$) associated with a
prescribed rate-profile $\mv{\alpha}$ under a LTPC $p^*$ is defined
as the maximum sum-rate of users over the expected (delay-limited)
capacity region under the constraint that the average (constant)
transmit rate of each user $r_k^*$ must satisfy
$\frac{r_i^*}{r_j^*}=\frac{\alpha_i}{\alpha_j}, \forall
i,j\in\{1,\ldots,K\}$.
\end{definition}

\begin{definition}\label{def:delay penalty}
For some given delay profile $\mv{\alpha}$ and LTPC $p^*$, the delay
penalty $C_{\rm DP}(p^*,\mv{\alpha})$ is equal to $C_{\rm
e}(p^*,\mv{\alpha})-C_{\rm d}(p^*,\mv{\alpha})$.
\end{definition}

\begin{figure}
\psfrag{a}{$R_1$}\psfrag{b}{$R_2$}\psfrag{c}{$\mathcal{C}_{\rm
e}(p^*)$}\psfrag{d}{$\mathcal{C}_{\rm
d}(p^*)$}\psfrag{e}{$\mv{\alpha}$}\psfrag{f}{$C_{\rm
DP}(p^*,\mv{\alpha})$}\psfrag{g}{$R_1+R_2=C_{\rm
d}(p^*,\mv{\alpha})$}\psfrag{h}{$R_1+R_2=C_{\rm
e}(p^*,\mv{\alpha})$}
\begin{center}
\scalebox{1.0}{\includegraphics*[41pt,600pt][300pt,761pt]{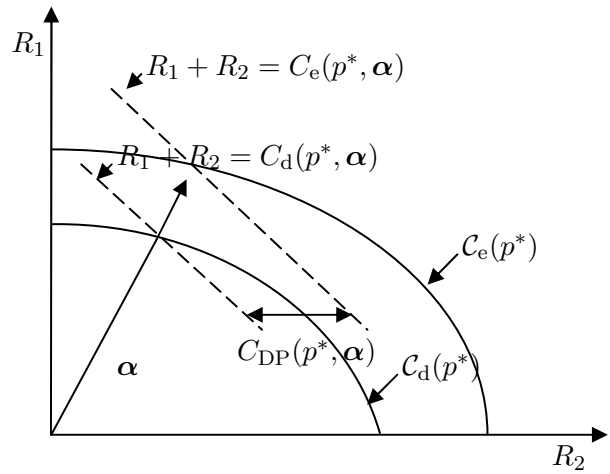}}
\caption{Illustration of the delay penalty $C_{\rm
DP}(p^*,\mv{\alpha})$. }\label{fig:delay penalty}
\end{center}
\end{figure}

The proposed definition for delay penalty is illustrated in Fig.
\ref{fig:delay penalty} for a 2-user case. From Definition
\ref{def:delay penalty}, it is noted that characterization of
$C_{\rm DP}(p^*,\mv{\alpha})$ requires that of both expected and
delay-limited throughput. In the next, we present the algorithm for
characterizing $C_{\rm e}(p^*,\mv{\alpha})$ for some given $p^*$ and
$\mv{\alpha}$. Similar algorithm can also be developed for
characterizing $C_{\rm d}(p^*,\mv{\alpha})$ and is thus omitted here
for brevity. According to Definition \ref{def:throughput} and
(\ref{eq:expected capacity region}) and using the BC-MAC duality
result in Section \ref{sec:system model}, the expected throughput
$C_{\rm e}(p^*,\mv{\alpha})$ can be obtained by considering the
following optimization problem ({\bf P4}):
\begin{eqnarray}
\mathtt{Maximize} &&  C_e \\
\mathtt {Subject \ to} && \mathbb{E}_n[R_k(n)] \geq C_e\alpha_k,  \\
&& \mv{R}(n)\in\mathcal{C}^{\rm MAC}_n\left(\{q_k(n)\},
\{\mv{h}_k^{\dag}(n)\}\right), \ \forall n
\\&& q_k(n)\geq 0 \ \forall n,k. \\ && \mathbb{E}_n\left[\sum_{k=1}^K
q_k(n)\right]\leq p^*
\end{eqnarray}
Similar like Problem P1, it can be verified that the above problem
is also convex, and hence can be solved using convex optimization
techniques. Here, instead of solving Problem P4 from a scratch, the
proposed solution transforms this problem into a special form of
Problem P1 and hence the same algorithm for Problem P1 can be
applied. First, considering the following transmit power
minimization problem ({\bf P5}) for support of any arbitrary set of
average-rate demands $\{R^*_k\}$ that satisfy a given rate-profile
constraint $\mv{\alpha}$, i.e.,
$\frac{R_i^*}{R_j^*}=\frac{\alpha_i}{\alpha_j}$, $\forall
i,j\in\{1,\ldots,K\}$:
\begin{eqnarray}
\mathtt{Minimize} && \mathbb{E}_n\left[\sum_{k=1}^K q_k(n)\right] \\
\mathtt {Subject \ to} && \mathbb{E}_n[R_k(n)] \geq R_{\rm sum}^*\alpha_k, \ \forall k, \\
&& \mv{R}(n)\in\mathcal{C}^{\rm MAC}_n\left(\{q_k(n)\},
\{\mv{h}_k^{\dag}(n)\}\right), \ \forall n
\\&& q_k(n)\geq 0 \ \forall n,k.
\end{eqnarray}
Note that $R_{\rm sum}^*\triangleq\sum_{k=1}^{K}R^*_k$. It is
observed that the above problem is a special case of Problem P1 if
all users have NDC transmission, i.e., those constant-rate
constraints for DC users in (\ref{eq:CR constraint}) are removed.
Hence, Problem P5 can also be solved by the proposed algorithm in
Table \ref{table:minpower algorithm}. Let $q^*$ denote the minimal
transmit power obtained after solving Problem P5. Notice that
$C_{\rm e}(p^*,\mv{\alpha})$ is a non-decreasing function of $p^*$
with some given $\mv{\alpha}$ because the expected capacity region
$\mathcal{C}_{\rm e}(p^*)$ corresponding to a larger $p^*$ always
contains that with a smaller $p^*$. Hence, if $p^*> q^*$, it can be
inferred that $C_{\rm e}(p^*,\mv{\alpha})$ must be larger than the
assumed $R_{\rm sum}^*$. Otherwise, $C_{\rm e}(p^*,\mv{\alpha})\leq
R_{\rm sum}^*$. By using this property, $C_{\rm e}(p^*,\mv{\alpha})$
can be easily obtained by a bisection search \cite{Boydbook}.

\begin{figure}
\psfrag{a}{$R_1$}\psfrag{b}{$R_2$}\psfrag{c}{$\mv{\alpha}_{\rm
e}$}\psfrag{d}{$\mv{\alpha}^*$}\psfrag{g}{
$R_1+R_2=C_e(p^*,\mv{\alpha}^*)$}\psfrag{h}{
$R_1+R_2=C_e(p^*,\mv{\alpha}_e)$}\psfrag{i}{$C_{\rm
FP}(p^*,\mv{\alpha}_{\rm e})$}
\begin{center}
\scalebox{1.0}{\includegraphics*[41pt,600pt][300pt,761pt]{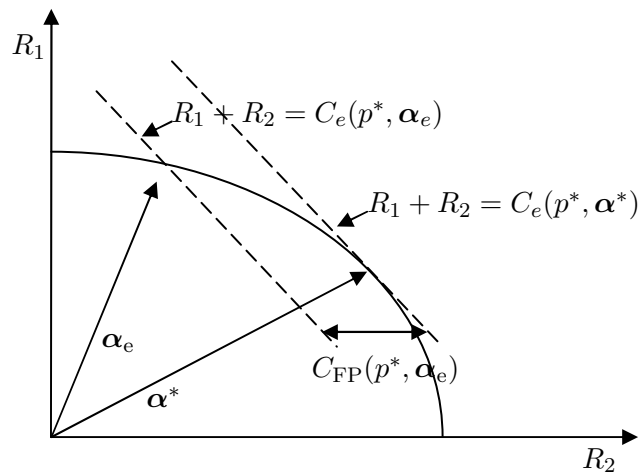}}
\caption{Illustration of the fairness penalty $C_{\rm
FP}(p^*,\mv{\alpha}_{\rm e})$ in the expected capacity region
$\mathcal{C}_e(p^*)$.}\label{fig:sum rate capacity}
\end{center}
\end{figure}

{\bf Fairness Penalty:} There is an interesting relationship between
the conventional sum capacity over the expected capacity region, and
the expected throughput as a function of delay-profile, as shown in
Fig. \ref{fig:sum rate capacity} for a 2-user case. The sum capacity
is obtained by maximizing the sum-rate of users over the expected
capacity region so as to maximally exploit the multiuser diversity
gain in the achievable network throughput. However, it does not
guarantee the resultant rate allocation among users. Let the
resultant rate portion allocated to each user in the sum capacity be
specified by a rate-profile vector $\mv{\alpha}^*$. In contrast, the
expected throughput maximizes the sum-rate of users under any
arbitrary rate-profile vector $\mv{\alpha}_{\rm e}$. Due to this
hard fairness constraint, the expected throughput in general is
smaller than the sum capacity if $\mv{\alpha}_{\rm e}$ is different
from $\mv{\alpha}^*$, and their difference can be used as a measure
of the {\it fairness penalty}, which is denoted by $\mathcal{C}_{\rm
FP}(p^*,\mv{\alpha}_{\rm e})\triangleq C_{\rm
e}(p^*,\mv{\alpha}^*)-C_{\rm e}(p^*,\mv{\alpha}_e)$. Similarly, the
fairness penalty can also be defined in the delay-limited case.

{\bf Asymptotic Results:} Consider the MISO-BC with asymptotically
large number of users $K$ but fixed number of transmit antennas $M$
at the BS. It is assumed that the network is homogeneous where all
users have mutually independent but identically distributed
channels, and have identical rate demands, i.e.,
$\alpha_k=\frac{1}{K}, \forall k$. Under these assumptions, in
\cite{Sharif07} it has been shown that as $K\rightarrow \infty$ the
expected throughput $C_{\rm e}$ under any finite power constraint
$p^*$ scales like $M\log_2 \log K+\mathcal{O}(1)$. In the following
theorem, we provide this asymptotic result for the delay-limited
throughput:
\begin{theorem}\label{theorem:DLC throughput}
Under the assumption of symmetric fading and symmetric user rate
demand, the delay-limited throughput $C_{\rm d}$ for a fading
MISO-BC under a LTPC $p^*$ is upper-bounded by
$\frac{p^*}{\rho\log2}$ as $K\rightarrow \infty$, where $\rho$ is a
constant depending solely on the channel distribution.
\end{theorem}
\begin{proof}
Please refer to Appendix \ref{appendix:proof theorem}.
\end{proof}
The above results suggest very different behaviors of the achievable
network throughput with NDC versus DC transmission as $K$ becomes
large. On the one hand, for the expected throughput, transmission
delay is not an issue and hence the optimal strategy is to select
only a subset of users with the best joint channel realizations for
transmission at one time. The expected throughput thus scales
linearly with $M$ and in double-logarithm with $K$, for which the
former is due to spatial multiplexing gain and the latter arises
from the multiuser-diversity gain. On the other hand, in the
delay-limited case, a constant-rate transmission needs to be ensured
for all users. Consequently, as the number of users increases,
though more degrees of freedom are available for optimizing transmit
parameters such as precoding vectors and encoding order of users,
the delay-limited throughput is eventually saturated. The above
comparison demonstrates that for a SDMA-based network with a large
user population, transmission delay can be a critical factor that
prevents from achieving the maximum asymptotic throughput. However,
notice that this may not be the case for the network having similar
$M$ and $K$, as will be verified later by the simulation results.

\section{Simulation Results} \label{sec:simulation results}

In this section, simulation results are presented for evaluating the
performances of dynamic resource allocation for the fading MISO-BC
under various transmission delay considerations. Since the network
throughput is contingent on transmit delay requirements as well as
many other factors such as number of transmit antennas at the BS,
number of mobile users, user channel conditions and rate
requirements, and the transmit power constraint at the BS, various
combinations of these factors are considered in the following
simulations with an aim to demonstrate how transmission delay
interplays with other factors in determining the achievable network
throughput. The simulation results are presented in the following
subsections. Note that in the following simulation, user channel
vectors $\{\mv{h}_k(n)\}$ are independently generated from the
population of CSCG vectors, and if not stated otherwise, it is
assumed that $\mv{h}_k(n)\sim\mathcal{CN}(\mv{0},\mv{I}), \forall
k$.

\subsection{Transmit Optimization for Mixed NDC and DC Traffic}

First, consider a MISO-BC with $M=K=4$ with two users having NDC
transmission and the other two having DC transmission. For
convenience, it is assumed that the target average transmit rates
for the two users with NDC transmission are both equal to $R_{\rm
NDC}^*$, and the target constant rates for the two users with DC
transmission both equal to $R_{\rm DC}^*$. Let $\gamma$ denote the
{\it loading factor} representing the ratio of the total amount of
NDC traffic to that of the sum of NDC and DC traffic, i.e.,
$\gamma\triangleq\frac{R_{\rm NDC}^*}{R_{\rm NDC}^*+R_{\rm DC}^*}$.
If the proposed algorithm in Table \ref{table:minpower algorithm}
that achieves optimal dynamic resource allocation is used, the
required average transmit power at the BS is expected to be the
minimum for satisfying both NDC and DC user rate demands for any
given $\gamma$. For purpose of comparison, two suboptimal
transmission schemes are also considered:
\begin{itemize}
\item {\it Time-Division-Multiple-Access (TDMA):} A simple transmission scheme is to
divide each transmission period into $K$ consecutive equal-duration
time-blocks, each dedicated for transmission of one user's data
traffic. If coherent precoding is applied at each fading state $n$,
i.e., the precoder for user $k$, defined as
$\mv{\hat{b}}_k(n)\triangleq\frac{\mv{b}_k(n)}{\|\mv{b}_k(n)\|}$, is
equal to  $\frac{\mv{h}_k^{\dag}(n)}{\|\mv{h}_k(n)\|}, \forall k$,
it can be shown that the MISO-BC is decomposable into $K$
single-user SISO channels. Depending on each user's delay
requirement, conventional water-filling power control
\cite{Goldsmith97} and channel-inversion power control
\cite{Caire00} can be applied over different fading states for users
with NDC and DC transmission, respectively, to achieve the minimum
average transmit power under the given user rate demand.

\item {\it Zero-Forcing (ZF) -Based SDMA:} Another possible transmission
scheme is also based on the SDMA principle, i.e., supporting all
user's transmission simultaneously by multi-antenna spatial
multiplexing at the BS. However, instead of using the
dirty-paper-coding (DPC)-based non-linear precoding assumed in this
paper, a simple ZF-based linear precoding is employed at the BS. The
ZF-based precoder $\mv{\hat{b}}_k(n)$ for user $k$ at each fading
state $n$  maximizes the user's own equivalent channel gain
$\|\mv{h}_{k}(n)\mv{\hat{b}}_k(n)\|^2$ subject to the constraint
that its associated co-channel interference must be completely
removed, i.e., $\mv{h}_{k'}(n)\mv{\hat{b}}_k(n)=0, \forall k'\neq
k$. Like TDMA, ZF-based precoding also decomposes the MISO-BC into
$K$ (assuming $K\leq M$) single-user SISO channels, and thereby
optimal single-user power control schemes can be applied.
\end{itemize}

In Fig. \ref{fig:mixed traffic high power} and Fig. \ref{fig:mixed
traffic low power}, these three schemes are compared for two cases
of network throughput, one corresponding to the sum-rate of users
$2R_{\rm NDC}^*+2R_{\rm NDC}^*=6$ bits/complex dimension, and one
corresponding to 2 bits/complex dimension. The required average
transmit power over 500 randomly generated channel realizations is
plotted versus the loading factor $\gamma$. First, in Fig.
\ref{fig:mixed traffic high power} it is observed that in the case
of high throughput (bandwidth-limited), ZF-based SDMA outperforms
TDMA because of its larger spectral efficiency by spatial
multiplexing. However, in the case of low throughput
(power-limited), in Fig. \ref{fig:mixed traffic low power} it is
observed that TDMA achieves better power efficiency than ZF-based
SDMA. This is because coherent precoding in TDMA provides more
diversity and array gains, which become more dominant over spatial
multiplexing gains at the power-limited regime. Secondly, it is
observed that in both cases of high and low throughput, the proposed
scheme always outperforms both TDMA and ZF-based SDMA given any
loading factor $\gamma$. This is because the proposed scheme
optimally balances the achievable spatial multiplexing, array, and
diversity gains for the fading MISO-BC. Thirdly, it is observed that
for all schemes, the required transmit power associated with some
$\gamma$, $\gamma<0.5$, is always larger than that with $1-\gamma$
(e.g., comparing $\gamma=0.1$ and $\gamma=0.9$), i.e., given the
same portion of data traffic in the total traffic, NDC traffic has a
better power efficiency than DC traffic. This is because NDC traffic
allows for more flexible dynamic resource allocation than DC traffic
and thus leads to a better power efficiency.

\subsection{Convergence of Online Algorithm}

\begin{figure}
\centering{
 \epsfxsize=4.2in
    \leavevmode{\epsfbox{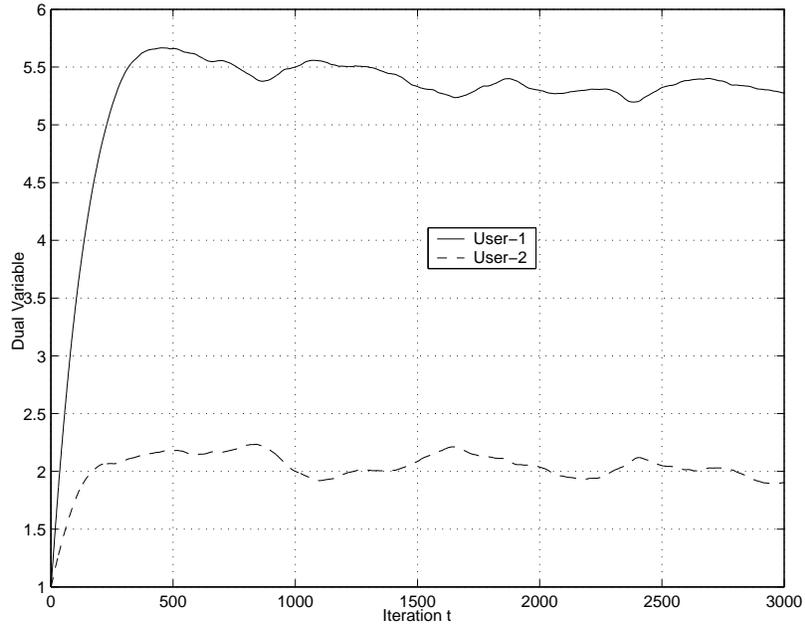}} }
\caption{The dual variable $\mu_k[t]$ at different time $t$ obtained
by the online algorithm.}\label{fig:mu online}
\end{figure}

The convergence of the online algorithm in Table \ref{table:online
algorithm} is validated by simulations. For simplicity, it is
assumed that only NDC users are present in the network. A MISO-BC
with $K=2$ and $M=4$ is considered. The target average-rates for
user-1 and user-2 are 3 and 1 bits/complex dimension, respectively.
The online algorithm is implemented for 3000 consecutive
transmissions with randomly generated channel realizations.
Initially, the dual variables are set as $\mu_1[0]=\mu_2[0]=1$, and
the estimated average transmit rates as
$\bar{R}_1[0]=\bar{R}_2[0]=0$. The updates for $\{\bar{R}_k[t]\}$
and $\{\mu_k[t]\}$ as $t$ proceeds are shown in Fig. \ref{fig:Rbar
online} and Fig. \ref{fig:mu online}, respectively, for
$\Delta=\epsilon=0.01$. It is observed that the online algorithm
converges to the optimal dual variables and target average-rates for
both users after a couple of hundreds of iterations. Simulations
results (not shown in this paper) for other values of $\Delta$ and
$\epsilon$ indicate that in general a larger step size leads to a
faster algorithm convergence but also results in more frequent
oscillations.

\subsection{Throughput-Delay Tradeoff}

\begin{figure}
\centering{
 \epsfxsize=4.2in
    \leavevmode{\epsfbox{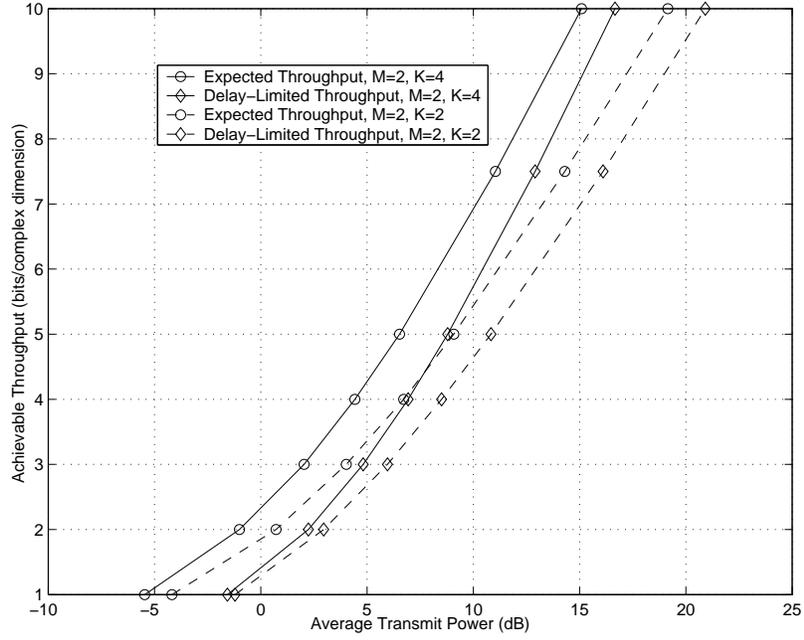}} }
\caption{Comparison of the expected throughput and the delay-limited
throughput.}\label{fig:comp DP}
\end{figure}

Fig. \ref{fig:comp DP} compares the network throughput under two
extreme cases of delay constraint considered in this paper, namely,
the expected throughput and the delay-limited throughput. It is
assumed that $M=2$, and two types of networks are considered: One is
a 2-user network with user rate profile $\mv{\alpha}=[\frac{2}{3} \
\frac{1}{3}]$; and the other is a 4-user network with
$\mv{\alpha}=[\frac{2}{6} \ \frac{2}{6} \ \frac{1}{6} \
\frac{1}{6}]$.  Notice that the ratio of rate demand between any of
the first two users and any of the last two users in the second case
is the same as that between the two users in the first case, which
is equal to $2$. First, it is observed that for both networks, the
delay penalties are only moderate for all considered transmit power
values. The delay penalty increases as $K$ becomes larger than $M$,
but only slightly. Small delay penalties in both cases can be
explained by extending the multi-antenna channel hardening effect
\cite{Tarokh04} in the single-user case to the fading MISO-BC, i.e.,
as the number of degrees of fading increases with $M$ for some
constant $K$, not only the mutual information associated with each
user's channel, but also the whole capacity region of the BC becomes
less variant over different fading states. Therefore, the sum-rate
of users for any given rate profile also changes less dramatically,
and as a consequence, imposing a set of strict constant-rate
constraints at each fading state (equivalent to a fixed
rate-profile) does not incur a large throughput loss for the fading
MISO-BC.

Secondly, it is observed that the expected throughput for a 4-user
network outperforms that for a 2-user network given that both
networks have the similar allocated rate portion among users. This
throughput gain can be explained by the well-known multiuser
diversity effect \cite{Tse02} for NDC transmission. As the number of
degrees of fading increases with $K$ for some constant $M$, the BS
can select relatively fixed number of users (around $M$) with the
best joint channel realizations from a larger number of total users
for transmission at each time. Thereby, the network throughput is
boosted provided that each user has sufficient delay tolerance. On
the other hand, it is observed that, maybe more surprisingly, the
delay-limited throughput for the 4-user network also outperforms
that for the 2-user network under the similar allocated rate portion
among users. Notice that in this case, all user's rates need to be
constant at each fading state, and consequently, adaptive rate
allocation for achieving the expected throughput does not apply
here. This throughput gain in the DC case can also be explained by a
more generally applied multiuser diversity effect. Even though a set
of constant-rates of users need to be satisfied at each fading state
in the DC case, a larger number of users provides the BS more
flexibility in jointly optimizing allocation of transmit resources
among users based on their channel realizations. This multiuser
diversity gain becomes more substantial as $K$ increases because a
new user can bring along additional $M$ degrees of fading. However,
it is also important to take note that the multiuser diversity gain
in the delay-limited case does have certain limitation, e.g., as $K$
becomes overwhelmingly larger than $M$, the delay-limited throughput
eventually gets saturated (see Theorem \ref{theorem:DLC
throughput}).

\subsection{Throughput-Fairness Tradeoff}
\begin{figure}
\centering{
 \epsfxsize=4.2in
    \leavevmode{\epsfbox{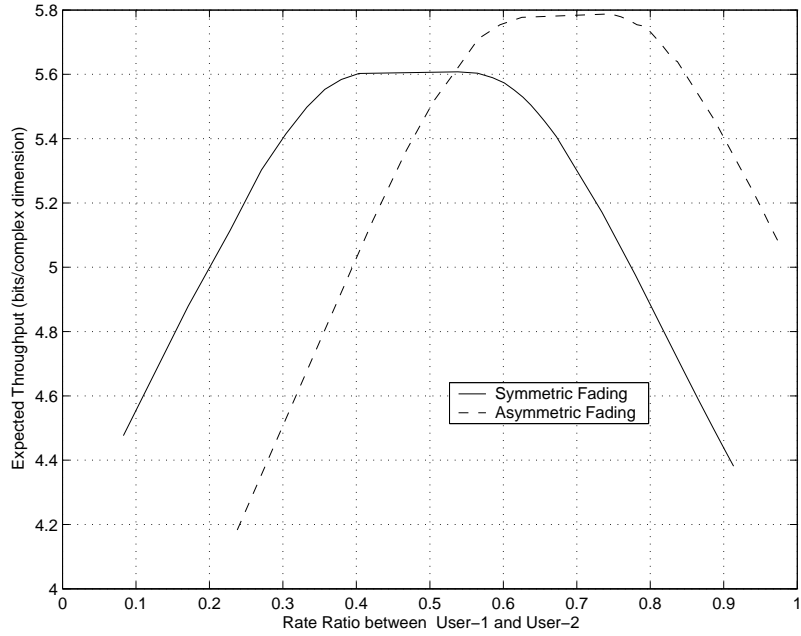}} }
\caption{Comparison of the expected throughput under different
fairness constraints.}\label{fig:comp FP}
\end{figure}

At last, the tradeoff between the achievable network throughput and
the fairness for user rate allocation is demonstrated. The NDC
transmission is considered and hence the expected throughput is of
interest. A network with 2 users is considered and it is assumed
that $M=2$, and the average LTPC at the BS is fixed as 10. Two
fading channel models are considered: One has symmetric fading where
both user channel vectors $\mv{h}_k(n), k=1,2,$ are assumed to be
distributed as $\mathcal{CN}(\mv{0},\mv{I})$; and the other has
asymmetric fading where
$\mv{h}_1(n)\sim\mathcal{CN}(\mv{0},2\mv{I})$, and
$\mv{h}_2(n)\sim\mathcal{CN}(\mv{0},\frac{1}{2}\mv{I})$. Notice that
the asymmetric-fading case may correspond to a near-far situation in
the cellular network where user-1 is closer to the BS and hence has
an average channel gain of approximately $20\times \log_{10}4=12$ dB
compared to user-2. Let $\phi$ denote the ratio of average-rate
demand between user-1 and user-2, i.e., for the corresponding rate
profile $\mv{\alpha}$, $\frac{\alpha_1}{\alpha_2}=\phi$. In Fig.
\ref{fig:comp FP}, the expected throughput is shown as a function of
$\phi$ for both symmetric- and asymmetric-fading cases. It is
observed that in the symmetric-fading case, a strict fairness
constraint for equal rate allocation among users, i.e., $\phi=0.5$,
also corresponds to the maximum expected throughput or the sum
capacity. In contrast, for the case of asymmetric fading, the
maximum expected throughput is achieved when $\phi=0.7$, i.e.,
user-1 is allocated 70\% of the expected throughput because of its
superior channel condition. However, in the latter case, a strict
fairness constraint with $\phi=0.5$ yields a throughput loss of only
0.3 bits/complex dimension. This small fairness penalty can be
explained by observing that the expected throughput for the fading
MISO-BC under both symmetric and asymmetric fading is quite
insensitive to $\phi$ at a very large range of its values,
indicating that by optimizing resource allocation, transmission with
very heterogeneous rate requirements may incur only a negligible
network throughput loss.

\section{Concluding Remarks}\label{sec:conclusion}

This paper investigates the capacity limits and the associated
optimal dynamic resource allocation schemes for the fading MISO-BC
under various transmission delay and fairness considerations. First,
this paper studies transmit optimization with mixed
delay-constrained and delay-tolerant data traffic. The proposed
online resource allocation algorithm is based on a two-layer
Lagrange-duality method, and is amenable to real-time cross-layer
implementation. Secondly, this paper characterizes some key
fundamental tradeoffs between network throughput, transmission delay
and user fairness in rate allocation, and draws some novel insights
pertinent to multiuser diversity and channel hardening effects for
the fading MISO-BC. This paper shows that when there are similar
numbers of users and transmit antennas at the BS, the delay penalty
and the fairness penalty in the achievable network throughput may be
only moderate, suggesting that employing multi-antennas at the BS is
an effective means for delivering data traffic with heterogeneous
delay and rate requirements.

The results obtained in this paper can also be extended to the
uplink transmission in a cellular network by considering the fading
MAC under individual user power constraint instead of the sum-power
constraint in this paper as a consequence of the BC-MAC duality.
Hence, another important factor needs to be taken into account by
dynamic resource allocation for the fading MAC is the fairness in
user transmit power consumption. The concept of rate profile in this
paper can also be applied to define a similar power profile for
regulating the power consumption between users for the MAC
\cite{Mohseni06}. Furthermore, although the developed results in
this paper are under the assumption of capacity-achieving
transmission using DPC-based non-linear precoding at the BS, they
are readily extendible to other suboptimal transmission methods such
as linear precoding provided that the achievable rate region by
these methods is still a convex set and, hence, like in this paper
similar convex optimization techniques can be applied.

\appendices
\section{Alternative Algorithm for Problem P3}
\label{appendix:algorithm}

In Problem P3, the constraints are separable and the objective
function is convex. Hence, this problem can be solved iteratively by
the block-coordinate decent method \cite{Bertsekasbook}. At each
iteration, this method minimizes the objective function w.r.t. one
$q_k$ while holding all the other $q_k$'s constant. More
specifically, the method minimizes (\ref{eq:objective func}) w.r.t.
$q_1$ with constant $\{q_2,\ldots,q_K\}$, and then $q_2$ with
constant $\{q_1,q_3,\ldots,q_K\},\ldots$, to $q_K$ with constant
$\{q_1,\ldots,q_{K-1}\}$, and after that the above routine is
repeated. Because after each iteration the objective function only
decreases, the convergence to the global minimum of the objective
function is ensured. Each iteration of the above algorithm is
described as follows. Without loss of generality, assuming that in
(\ref{eq:objective func}), $\pi(k)=k,k=1,\ldots,K$. Considering any
arbitrary iteration for minimizing (\ref{eq:objective func}) w.r.t.
$q_m$, $m\in\{1,\ldots,K\}$, with all the other $q_k$'s constant,
Problem P3 can be rewritten as
\begin{eqnarray}
\mathtt{Minimize} && q_m- \sum_{k=m}^{K}
\left(\beta_k-\beta_{k+1}\right)\log_2\left|\mv{h}_{m}^{\dag}\mv{h}_{m}q_{m}+\sum_{i=1,i\neq
m}^{k}\mv{h}_{i}^{\dag}\mv{h}_{i}q_{i} +\mv{I}\right| \\ \mathtt
{Subject \ to} && q_m\geq 0.
\end{eqnarray}
By introducing the dual variable $\lambda_m$,  $\lambda_m\geq 0$,
associated with the constraint $q_m\geq 0$, the Karush-Kuhn-Tacker
(KKT) optimality conditions \cite{Boydbook} state that the optimal
$p_m^*$ and dual variable $\lambda_m^*$ for this problem must
satisfy
\begin{eqnarray}
\sum_{k=m}^{K}
\left(\beta_k-\beta_{k+1}\right)\mv{h}_{m}\left(\mv{h}_{m}^{\dag}\mv{h}_{m}q_{m}^*+\sum_{i=1,i\neq
m}^{k}\mv{h}_{i}^{\dag}\mv{h}_{i}q_{i} +\mv{I}\right)^{-1}\mv{h}_{m}^{\dag}&=&(1-\lambda_m^*)\log2, \label{eq:KKT1}\\
\lambda_m^*q_m^*&=&0. \label{eq:KKT2}
\end{eqnarray}
Let $d(q_m^*)$ denote the function on the left-hand-side of
(\ref{eq:KKT1}). From (\ref{eq:KKT2}), it is inferred that $q_m^*>0$
only if $\lambda_m^*=0$. From (\ref{eq:KKT1}) and by taking note
that $d(q_m^*)$ is a non-increasing function of $q_m^*$ for
$q_m^*\geq 0$, it follows that $q_m^*>0$ occurs only if
$d(0)>\log2$. Thus, it follows that
\begin{eqnarray}
q_m^*=\left\{\begin{array}{ll} 0 & \ {\rm if} \ d(0)\leq \log2 \\
q_0 & {\rm otherwise,}
\end{array}\right.
\end{eqnarray}
where $q_0$ is the unique root for $d(q_m^*)=\log2$. $q_0$ can be
easily found by a bisection search \cite{Boydbook} over
$[0,\frac{\beta_m}{\log2}]$, where the above upper-bound for $q_0$
is obtained by the following inequalities and equality:
\begin{eqnarray}
\log2&\leq&\sum_{k=m}^{K}
\left(\beta_k-\beta_{k+1}\right)\mv{h}_{m}\left(\mv{h}_{m}^{\dag}\mv{h}_{m}q_{m}^*+\mv{I}\right)^{-1}\mv{h}_{m}^{\dag}
\label{eq:ineq 1}\\ &\leq& \frac{1}{q_{m}^*}\sum_{k=m}^{K}
\left(\beta_k-\beta_{k+1}\right) \label{eq:ineq 2} \\ &=&
\frac{\beta_m}{q_m^*}.
\end{eqnarray}

\section{Proof of Theorem \ref{theorem:DLC throughput}}
\label{appendix:proof theorem}

First, we obtain an upper-bound for the delay-limited throughput
$C_{\rm d}$ by assuming that there is no co-channel interference
between users, as opposed to the successive interference
pre-subtraction by DPC. Under this assumption, the fading MISO-BC is
decomposed into $K$ parallel single-user fading channels. Let
$\hat{p}_k$ denote the average transmit power assigned to user $k$,
$k=1,\ldots,K$. The maximum constant-rate achievable over all fading
states (or the so-called delay-limited capacity \cite{Tse98b}) for
user $k$ can be expressed as \cite{Caire00}
\begin{equation}
C_{\rm d}(k)=\log_2\left(1+\frac{\hat{p}_k}{\rho_k}\right),
\end{equation}
where $\rho_k=\mathbb{E}_n\left[\frac{1}{\|\mv{h}_k(n)\|^2}\right]$.
Because of the assumed symmetric fading (hence, $\rho_k=\rho,
\forall k$) and symmetric rate demand (hence, $C_{\rm
d}(k)=\frac{C_{\rm d}}{K}, \forall k$), it follows that
$\hat{p}_k=\frac{p^*}{K}, \forall k$, achieves the maximum average
sum-rate of users. Hence, the delay-limited throughput is
upper-bounded by
\begin{equation}\label{eq:sum DLC}
C_{\rm d}\leq K\log_2\left(1+\frac{p^*}{K\rho}\right),
\end{equation}
which holds for any $K\geq 1$. By taking the limit of the
right-hand-side of (\ref{eq:sum DLC}) as $K\rightarrow \infty$, the
proof of Theorem \ref{theorem:DLC throughput} is completed.

\newpage

\begin{table}
\begin{center}
\begin{tabular}{|l|}
\hline

\hspace*{0.0cm} Initialize $\{\mu_k\}, k\in\mathcal{U}_{\rm NDC}$ and $\{\delta_k(n)\}, k\in\mathcal{U}_{\rm DC}$\\

\hspace*{0.0cm} While not all $\mathbb{E}_n[R'_k(n)]$ converges to
$R_k^*$, $k\in\mathcal{U}_{\rm
NDC}$, do \\

\hspace*{0.5cm} For $n=1,2,\ldots$  do \\

\hspace*{1.0cm} While not all $R'_k(n)$ converges to $R_k^*$,
$k\in\mathcal{U}_{\rm DC}$, do \\

\hspace*{1.5cm} Solve Problem P3 for fading state
$n$ to obtain $\{R'_k(n)\}$\\

\hspace*{1.5cm} Update $\{\delta_k(n)\}$ according to (\ref{eq:delta
update}). \\

\hspace*{1.0cm} End While \\

\hspace*{0.5cm} End For \\

\hspace*{0.5cm} Update $\{\mu_k\}$ according to (\ref{eq:mu
update}). \\

\hspace*{0.0cm} End While \\

\hline
\end{tabular}
\end{center}
\caption{Proposed Algorithm For Problem P1.} \label{table:minpower
algorithm}
\end{table}

\begin{table}
\begin{center}
\begin{tabular}{|l|}
\hline

\hspace*{0.0cm} Initialize $\{\mu_k[0]\}, k\in\mathcal{U}_{\rm NDC}$.  \\

\hspace*{0.0cm} Set $\bar{R}_k[0]=0, \forall k\in \mathcal{U}_{\rm NDC}$; $t=1$.\\

\hspace*{0.0cm} Repeat \\

\hspace*{0.5cm} $\mu_k[t]\leftarrow
\left\{\mu_k[t-1]+\Delta\left(R_k^* -
\bar{R}_k[t-1]\right)\right\}^+,
k\in\mathcal{U}_{\rm NDC}$. \\

\hspace*{0.5cm}  Initialize $\{\delta_k[t]\},k\in\mathcal{U}_{\rm
DC}$.\\

\hspace*{0.5cm} While not all $R_k[t]$ converges to $R_k^*$,
$k\in\mathcal{U}_{\rm DC}$, do \\

\hspace*{1.0cm} Solve Problem P3 for given $\{\mu_k[t]\}$, $\{\delta_k[t]\}$ and $\{\mv{h}_k^{\dag}[t]\}$ to obtain $\{R_k[t]\}$.\\

\hspace*{1.0cm} $\delta_k[t]\leftarrow
\left\{\delta_k[t]+\Delta\left(R_k^* - R_k[t]\right)\right\}^+,
k\in\mathcal{U}_{\rm DC}$. \\

\hspace*{0.5cm} End While \\

\hspace*{0.5cm} $\bar{R}_k[t]=(1-\epsilon)\bar{R}_k[t-1]+\epsilon
R_k[t]$. \\

\hspace*{0.5cm} $t\leftarrow t+1$. \\

\hline
\end{tabular}
\end{center}
\caption{An Online Algorithm For Problem P1.} \label{table:online
algorithm}
\end{table}

\begin{figure}
\centering{
 \epsfxsize=4.2in
    \leavevmode{\epsfbox{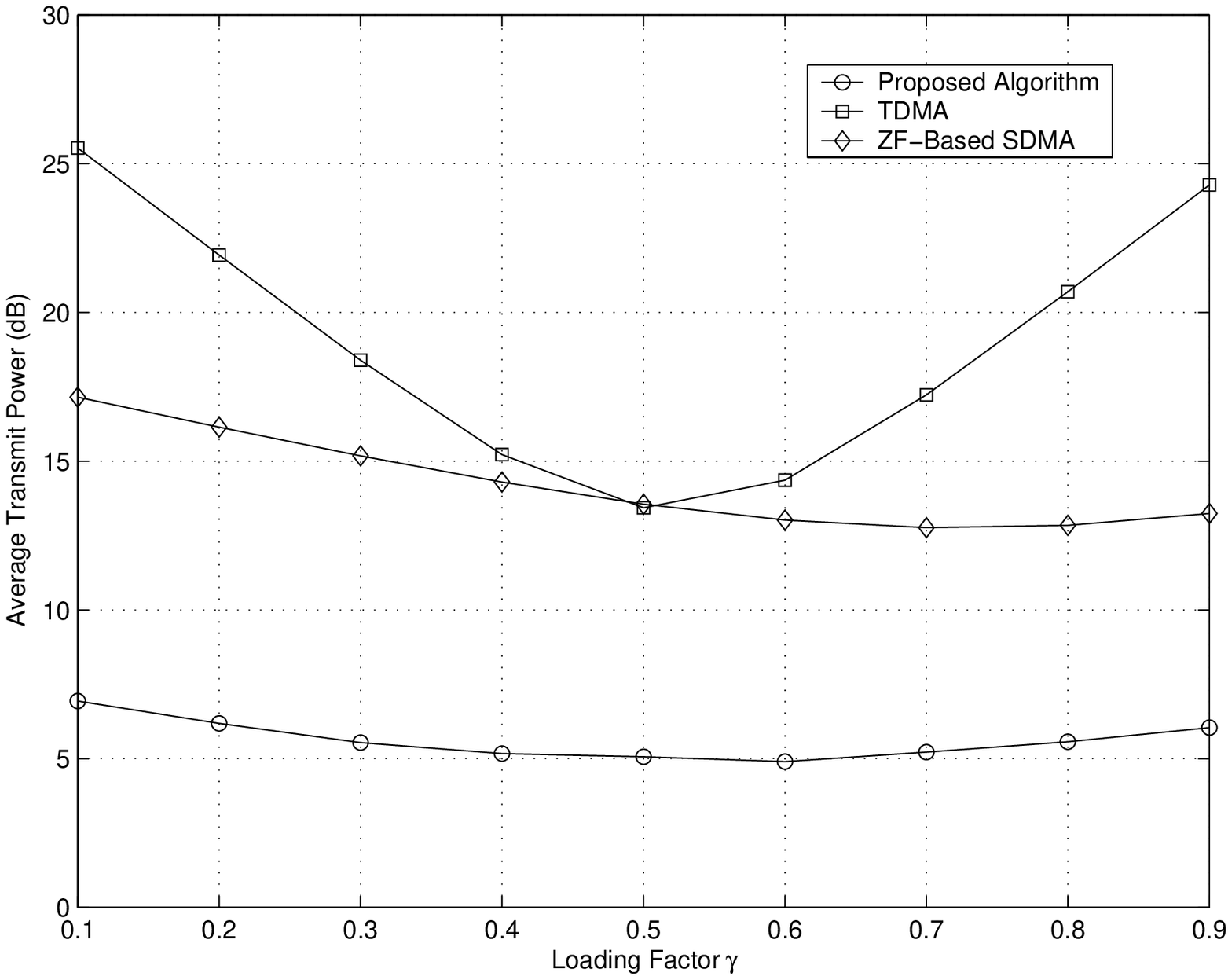}} }
\caption{Comparison of the average transmit powers for different
schemes under mixed NDC and DC transmission. The total amount of NDC
and DC traffic is 6 bits/complex dimension.}\label{fig:mixed traffic
high power}
\end{figure}

\begin{figure}
\centering{
 \epsfxsize=4.2in
    \leavevmode{\epsfbox{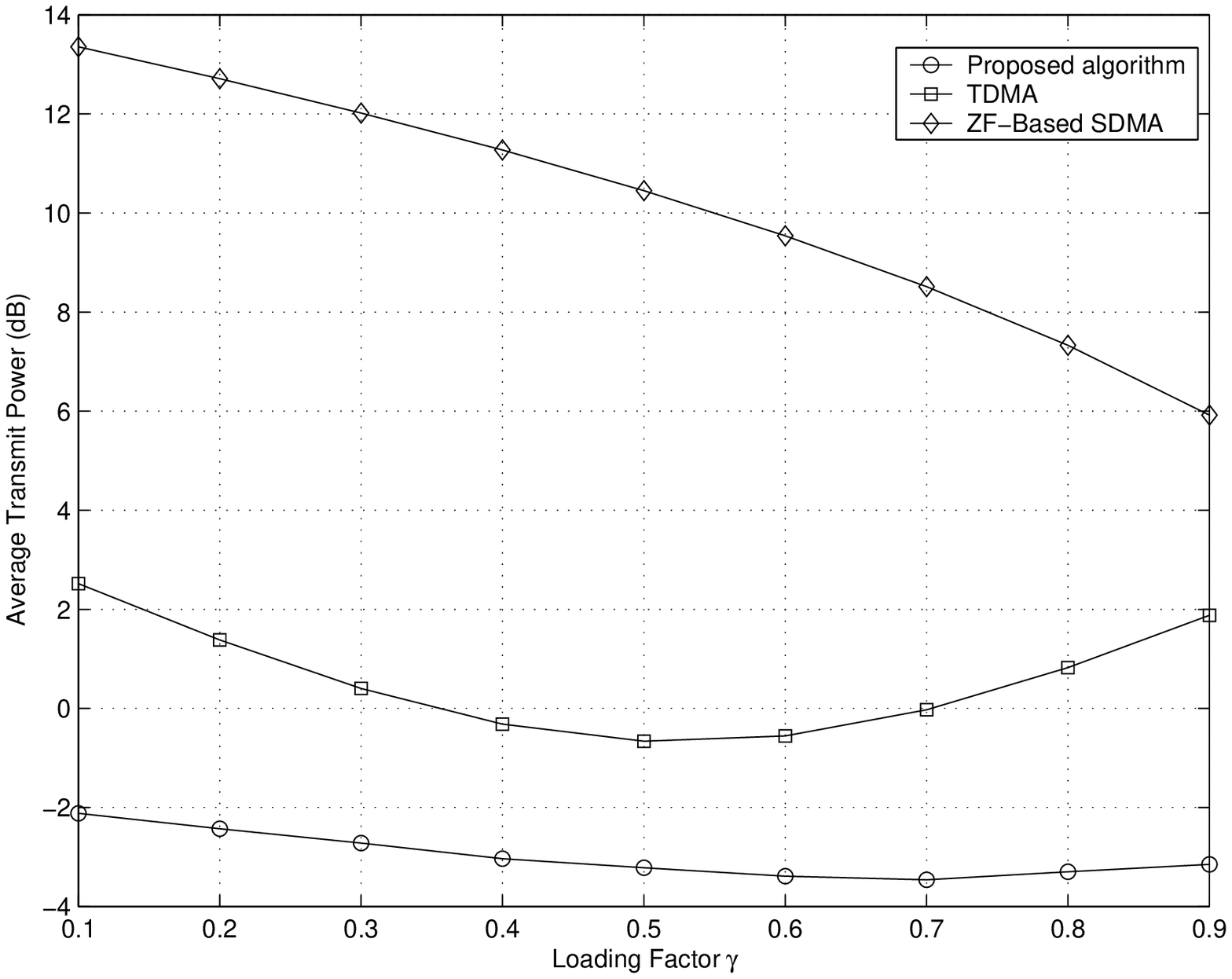}} }
\caption{Comparison of the average transmit powers for different
schemes under mixed NDC and DC transmission. The total amount of NDC
and DC traffic is 2 bits/complex dimension.}\label{fig:mixed traffic
low power}
\end{figure}

\begin{figure}
\centering{
 \epsfxsize=4.2in
    \leavevmode{\epsfbox{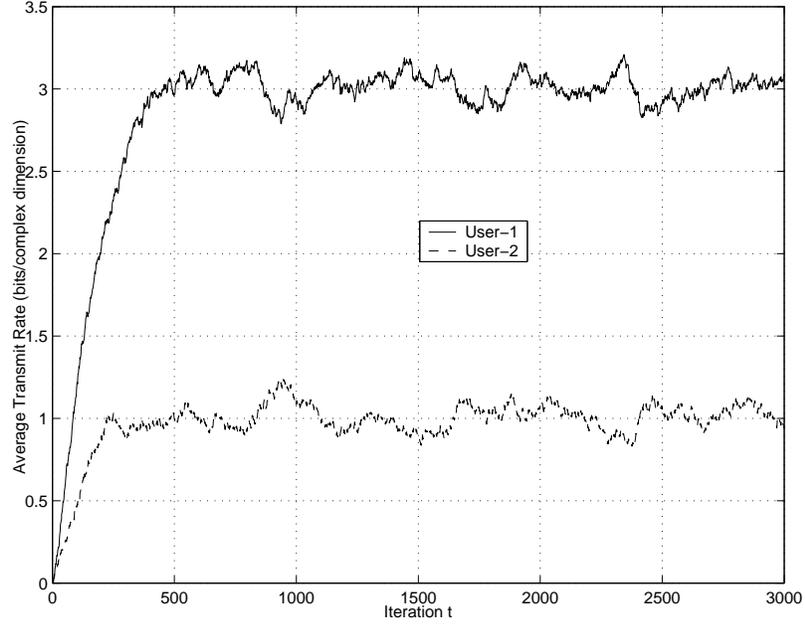}} }
\caption{The estimated average transmit rate $\bar{R}_k[t]$ at
different time $t$ obtained by the online algorithm. The target
average-rate for user-1 and user-2 are $3$ and $1$ bits/complex
dimension, respectively.}\label{fig:Rbar online}
\end{figure}

\end{document}